\documentclass[aps,prb,twocolumn,showpacs,10pt,superscriptaddress,floatfix,longbibliograpgy]{revtex4-2}

\usepackage[usenames, dvipsnames]{color}
\usepackage[normalem]{ulem}
\usepackage{amsmath}
\usepackage{float}
\usepackage{color}
\usepackage{bbold}
\usepackage{amsmath}
\usepackage{amsfonts}
\usepackage[edges]{forest}
\usepackage{psfrag}
\usepackage{rotating}

\usepackage[normalem]{ulem}
\usepackage{soul}
\usepackage{amssymb}
\usepackage{graphicx}
\usepackage{comment}

\usepackage[colorlinks, breaklinks, 
linkcolor=OrangeRed,
citecolor=RoyalBlue,
urlcolor=RoyalBlue]{hyperref}          
\usepackage{multirow}
\usepackage[all]{hypcap} 
\usepackage{xspace}
\usepackage{amssymb}
\usepackage{appendix}
\usepackage{pifont}

\usepackage{etoolbox} 
\usepackage{lipsum} 
\usepackage[capitalize]{cleveref}

\makeatletter
\appto{\appendix}{%
  \@ifstar{\def\theequation@prefix{A.}}%
          {}%
}
\makeatother

\begin{document}

\title{Fingerprint of Non-Hermiticity in $d$-wave Altermagnet}

\author{Gaurab Kumar Dash}
\affiliation{Department of Physics, Indian Institute of Technology Delhi, Hauz Khas 110016, New Delhi, India}

\author{Subhasis Panda}
\email{subhasis@phy.nits.ac.in}
\affiliation{Department of Physics, National Institute of Technology Silchar, Assam 788010, India}

\author{Snehasish Nandy}
\email{snehasish@phy.nits.ac.in}
\affiliation{Department of Physics, National Institute of Technology Silchar, Assam 788010, India}

\begin{abstract}
We develop the non-Hermitian counterpart of a new class of collinear magnets, dubbed altermagnets, delineated by net zero magnetization with momentum-dependent spin-splitting bands. \textcolor{black}{The application of an imaginary gauge field in a two-dimensional $d$-wave altermagnet injects a non-reciprocal intercell hopping without hosting exceptional points (EPs). This non-Hermitian phase hosts a point gap which remains robust in the presence and absence of Rashba spin-orbit coupling.} By attaching a ferromagnetic lead with the altermagnet, we uncover the emergence of two pairs of second-order EPs. We establish that each of the EPs is associated with a half-integer quantized topological charge, a hallmark signature of the non-Hermitian topology. The existence of this non-Hermitian exceptional phase has been further confirmed by linear variation with the respective momentum and coalescence of the spin expectation value at the EPs. Finally, we demonstrate that the application of a planar magnetic field to the junction not only tunes the location of the EPs but also can annihilate a single pair or even all pairs of EPs with opposite topological charges depending upon the field strength and direction.
\end{abstract}
\maketitle

\section{Introduction} Recently, a new subclass
of magnetic materials, coined as \textit{altermagnet (AM)}, have been emerging
as an exciting research landscape due to exhibiting a novel form of collinear magnetism beyond conventional ferro- and antiferromagnetic ordering~\cite{Kusunose_2019, Zunger_2020, Libor_2021, Jungwirth_2022, Tomas_2022}. Altermagnets are characterized by two distinct features: (i) \textit{no net magnetization} even in the absence of time-reversal symmetry (TRS), and (ii) \textit{momentum-dependent spin-splitting} in electronic bands without the help of spin-orbit coupling~\cite{Jungwirth_2022, Tomas_2022, Igor_2022, Sinova_2022, Spaldin_2024}. The net magnetization in AMs vanishes due to the alternating order of the magnetic moments in both the direct- and momentum space, making them distinct from ferromagnets, which also break the TRS. On the other hand, the non-relativistic spin splitting in AM arises because the sublattices of opposite spin in AM are connected by a proper or improper rotation and/or reflections~\cite{Jungwirth_2022, Tomas_2022, Spaldin_2024}. This is in sharp contrast to the conventional antiferromagnets where the sublattices are mapped onto each other by inversion or translation, manifesting a spin-degeneracy of the electronic bands~\cite{Blundell_2001}. These unique features of altermagnet not only foster a plethora of intriguing properties~\cite{Song_2022, Kriegner_2023, Bai_2023, Tsymbal_2023, Yan_2023, Jacob_2023, Yugui_2024, Zhang_2024, Papaj_2023, Ghorashi_2023, Cheng_2023, Jeffrey_2024, Hariki_2024, Antonenko_2024, Samokhin_2024} which are of fundamental interests but also paves the way for potential technological application ranging from spintronics to quantum information processing~\cite{Yao_2024}.

 Besides altermagnetism, the unique spectral properties and topological features of non-Hermitian (NH) systems have garnered significant attention in recent years in diverse areas such as optical and photonic systems\cite{sp_apl24,sp_natmat2019,sp_nanoph21,PhysRevB.107.165120}, quantum computing \cite{sp_quant-arxiv,sp_PRA}, non-reciprocal light propagation \cite{sp_nature24,sp_pra_24}, and precision measurements \cite{sp_prl23}. Of these, NH topological systems are at the forefront of research in condensed matter physics due to their ability to reveal new topological phases and unique properties like exceptional points (EP) and skin effects. EPs are branch-point singularities on the NH eigenvalue manifolds, where not only the eigenvalues degenerately merge but also their corresponding eigenvectors coalesce. From a more technical perspective, the eigenvalue's geometric multiplicity is less than its algebraic multiplicity at EP, thereby making the Hamiltonian defective \cite{sp_Heiss_2008}. Manipulating the EPs in the NH systems has become a rich and fertile ground for exploration. For example, near EP, the degenerate modes split up sub-linearly against a small perturbation, which signifies enhanced measurement sensitivity, leading to an EP-enhanced sensor design \cite{sp_prl_epsensor,sp_sciadv23,sp_optical_nature17}. 
 
Recently, it has been proposed that material junctions pave the way for the experimental realization of the EPs. For example, the junction of $s-$wave superconductor-ferromagnetic (FM) lead hosts odd-frequency pairing \cite{sp_prb_scfm22}, non-Hermitian Weyl physics in a three-dimensional topological insulator coupled to an FM lead \cite{sp_PRR_tifm}, customizable EPs in a junction of Rashba semiconductor-FM lead \cite{sp_rsfm}. Additionally, the Zeeman field provides a handle for further tuning the EPs \cite{sp_prbcayao24,PhysRevB.109.035418}. Therefore, the confluence of material junctions and non-Hermiticy gives rise to novel phenomena that augment the topological band theory landscape. In view of the current surge in research in the domain of non-Hermitian systems and altermagnetism, the interplay between them remains to be explored and, therefore, needs urgent attention.

\begin{figure*}
    \centering
    \includegraphics[width=1\linewidth]{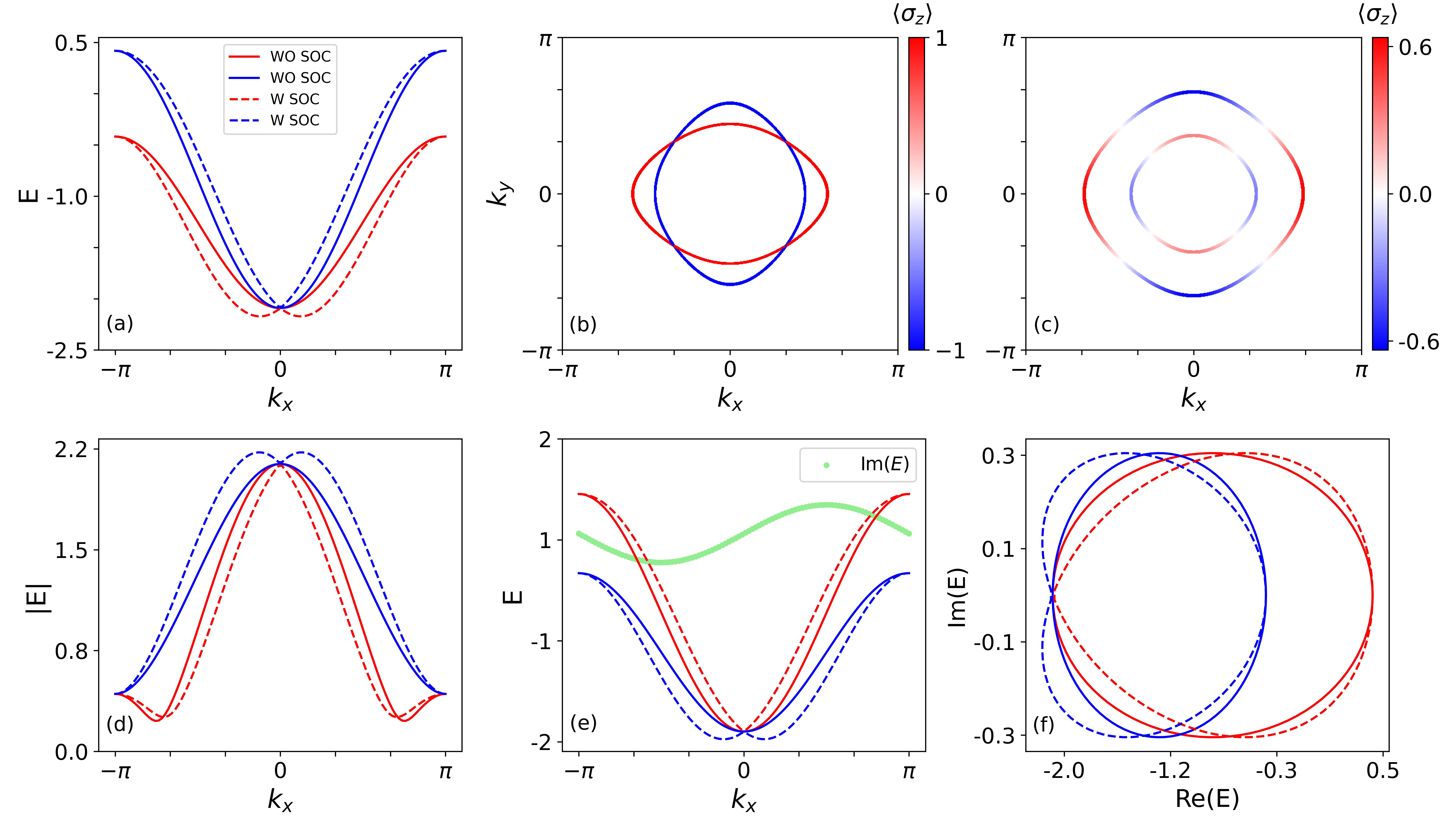}
    \caption{(a) shows the band dispersion along $k_x$ direction of Hermitian AM ($\mathcal{H}^{\rm H}_{\rm AM} (\mathbf{k})$) with $k_y = 0$ in the absence of RSOC, $\lambda = 0$ (solid lines) and in the presence of RSOC $\lambda = 0.4 \, t$ (dashed-lines). The red and blue colors represent valence and conduction bands, respectively. (b) and (c) depict Fermi surfaces at a chemical potential $\mu = -t$ without RSOC and with RSOC ($\lambda = 0.4 \, t$), respectively. The color code represents  $\langle \sigma_z \rangle$. (d) shows the absolute energy $\lvert E\rvert$ for non-Hermitian AM: $\mathcal{H}^{\rm NH}_{\rm AM} (\mathbf{k})$ versus $k_x$ with $k_{y}=0$ in the absence (solid lines) and presence (dashed-lines) of RSOC. The color codes and parameters are the same as (a). \textcolor{black}{(e) depicts the real (Re($E$)) and imaginary (Im($E$)) part
of the energy with respect to $k_x$ for $\lambda = 0$ (solid lines) and $\lambda = 0.4 \, t$ (dashed-lines). The Im($E$) is same for all scenario and is shown in green dotted-line. The complex energy eigenspectrum is displayed in (f) where $k_{y}$ is fixed to 0. \textcolor{black}{The other parameters are taken as $\tilde{t} = 0.5$, $\phi_I=0.3$, $t=2\tilde{t}\cosh \phi_I$, $\Delta = 0.2 \, t$, and $\delta = 2\tilde{t}\sinh \phi_I$.}}}
      \label{fig:dis}
\end{figure*}

In this Letter, we investigate the non-Hermitian physics in planar $d$-wave altermagnets. \textcolor{black}{In the presence of an imaginary gauge field, the system experiences intercell non-reciprocal hopping in both $x$ and $y$ directions and hosts point gap in the presence and absence of Rashba spin-orbit coupling without hosting EPs. We find that in the NH regime, the real part of the energy remains unaltered while the eigenspectrum is closed due to the asymmetric nature of the imaginary part of the energy.} Interestingly, by attaching an FM lead with AM, we propose the emergence of two pairs of second-order exceptional points associated with $\pm 1/2$ topological charge, a trademark signature of the NH topological physics. We highlight another unique signature of EP, demonstrating the linear parallel momentum dependence and coalescence of spin expectation at that point. Further, we show that the location and number of the emergent EPs can be tuned by an in-plane external magnetic field. Remarkably, by rotating the field direction, we find that the number of EPs can be reduced to a single pair or even null for a critical field strength. The recent evidence for altermagnetism reported in various materials (e.g., RuO$_2$, CrSb, MnTe) provides a platform to experimentally verify our predictions~\cite{Elmers_2024, Sato_2024, Zeng_2024, Yao_2024, Krempasky_2024}.

\section{Altermagnet with Imaginary Gauge Field} We consider a non-Hermitian two-dimensional single orbital model characterizing d-wave AM on a square lattice (lattice constant $a=1$), which can be written as~\cite{Yan_2023, Antonenko_2024, Samokhin_2024}
\begin{align}
\mathcal{H}^{\rm NH}_{\rm AM} (\mathbf{k}) &=\textcolor{black}{ -[t(\cos{k_x} + \cos{k_y}) - i\delta(\sin{k_x} + \sin{k_y})]\sigma_{0}} \nonumber \\
&+ \lambda(\sin{k_y}\sigma_x - \sin{k_x}\sigma_y) 
+ \Delta(\cos{k_x}-\cos{k_y})\sigma_z, \nonumber \\
&=\mathcal{H}^{\rm H}_{\rm AM} (\mathbf{k})+\textcolor{black}{i\delta (\sin{k_x} + \sin{k_y})\sigma_{0}},
\label{NH_Ham}
\end{align}
where $t$ parameterizes the usual kinetic energy, and $\Delta$ denotes d$_{x^{2}-y^{2}}$ altermagnetic order parameter. Here, $\sigma$ matrices act on the spin degree of freedom. $\lambda$ represents the strength of Rashba spin-orbit coupling (RSOC). \textcolor{black}{Here, we have introduced an imaginary gauge field $\phi_I$ in the kinetic part of the Hermitian Hamiltonian. Due to $\phi_I$, the nearest-neighbor hopping ($\tilde{t}$) along the positive (negative) $x$ and $y$ directions acquire a Peierl’s phase and are modified as $\tilde{t}e^{-\phi_I}$ ($\tilde{t}e^{\phi_I}$). \textcolor{black}{So, the kinetic part of the Hamiltonian can be written as
$-t(\cos k_x + \cos k_y)\sigma_0+i\delta(\sin k_x + \sin k_y)\sigma_0$ where $t=2\tilde{t}\cosh \phi_I$ and $\delta=2\tilde{t} \sinh \phi_I$.} This phase can be realized in experiments using coupled microring resonators in photonic systems~\cite{Hafezi_2013, Mittal_2014, Longhi, Longhi_2015, Longhi_2018}, cold atoms~\cite{Galitski} and electrical circuits~\cite{Huerta-Morales}.} 

In the context of symmetries, the Hermitian part of the above Hamiltonian breaks TRS ($\hat{\mathcal{T}}=-i\hat{\sigma}_{y} \hat{\mathcal{K}}$ where $\hat{\mathcal{K}}$ is the complex conjugation) and four-fold rotational symmetry about the $z$-axis ($\hat{C}_{4z} = e^{i\frac{\pi}{4} \hat{\sigma}_{z}}$), but remains invariant under the combination of $\hat{C}_{4z}\hat{\mathcal{T}}$. The RSOC breaks inversion symmetry without breaking $\hat{C}_{4z}\hat{\mathcal{T}}$. On the other hand, the non-Hermitian term breaks $\hat{C}_{4z}\hat{\mathcal{T}}$ symmetry. After doing the symmetry analysis, we move on to investigate the band structure and the associated Fermi surface of the altermagnet. In the Hermitian regime \textcolor{black}{($\delta = 0 ~ i.e., ~ \phi_I=0)$}, the band dispersion along the $k_x$ direction in the absence as well as in the presence of RSOC is shown in fig.~\ref{fig:dis}(a). We find anisotropic spin-polarized Fermi surface with diagonal degeneracies ($k_x=\pm k_y$) for $\lambda =0$ due to broken four-fold rotational symmetry by the altermagnetic order as can be seen in fig.~\ref{fig:dis}(b). On the other hand, it is clear from fig.~\ref{fig:dis}(c) that after turning on RSOC, the degeneracies along the diagonal direction are lifted, and spin-momentum locking appears on the spin-polarized Fermi surface.

By turning on the imaginary gauge field, the system enters into the NH regime ($\delta \neq 0$). The absolute energy dispersion along $k_x$-direction with $k_y=0$ is depicted in fig.~\ref{fig:dis}(d). It is clear from the figure that in addition to the degenerate point at $(0,0)$ of the Hermitian case, the gap also closes at $(\pi, 0)$. However, these points are not exceptional points because the non-Hermitian
term in eqn.~(\ref{NH_Ham}) contains an identity matrix $\sigma_0$ that does not affect the wavefunctions of the Hermitian system. Therefore, the wavefunctions of the eqn.~(\ref{NH_Ham})
remain orthonormal even at these degenerate points. The real and imaginary part of the energy (${\rm Re}(E)$ and Im($E$)) with respect to $k_x$ is shown in fig.~\ref{fig:dis}(e). It is clear that the real part of the spectrum remains the same for non-Hermitian bulk AM (as also evident from eqn.~(\ref{NH_Ham})). Additionally, the system experiences intercell non-reciprocal hopping in both directions and consequently, the imaginary part of the energy becomes momentum-dependent. The non-Hermitian term shifts the eigenenergy along the imaginary axis on the complex plane, producing a global growth or decay of eigenmodes. Moreover, we find that the real part is symmetric with respect to $k_x$, while the imaginary part exhibits an asymmetric property. Fig.~\ref{fig:dis}(f) portrays the complex energy eigenspectrum (${\rm Im}(E)$ vs Re($E$)). We fix one of the momenta, say $k_y=0$, making the NH Hamiltonian effectively one-dimensional. Remarkably, we find that the NH phase hosts point gap in the complex eigenspectrum irrespective of RSOC. The closed eigenspectrum validates the asymmetric nature of the imaginary part of the energy. This further hints the possibility of having skin effect in this system.

\section{Altermagnet-Ferromagnet Junction} To realize the exceptional signatures, we consider an open quantum system of altermagnet-ferromagnet (AM-FM) junction where a two-dimensional AM is coupled to the semi-infinite FM lead along $z \leq 0$. Due to half-space geometry, the coupling at the interface is designed by the spin-independent hopping between the last site of the FM lead ($z = 0$) and the first site of the AM at $z = 1$. The low-energy effective Hamiltonian describing the above system is given by
\begin{equation}
\mathcal{H}_{\rm AM-FM} (\mathbf{k}) = \mathcal{H}^{L}_{\rm AM} + \Sigma^{R}(\omega = 0),
\end{equation}
where $\mathcal{H}^{L}_{AM}$ is the low-energy part of $\mathcal{H}^{\rm H}_{\rm AM} (\mathbf{k})$ given in Eq.~(\ref{NH_Ham}) and $\Sigma^{R}(\omega = 0) = -i\xi^{+}\sigma_0 -i\xi^{-}\sigma_z$ represents the zero-frequency retarded spin-dependent self-energy (validated in the wide-band limit) engendering the coupling between the FM lead and AM. Here, $\xi^{\pm} = \frac{1}{2}(\xi_{\uparrow} \pm \xi_{\downarrow})$ with $\xi_{\sigma} = \pi\lvert t^{\prime}\rvert^2\rho^{FM}_{\sigma}$ and $\sigma = (\uparrow, \downarrow)$. The hopping amplitude from the AM to the FM lead is given by $t^{\prime}$ and $\rho^{FM}_{\sigma}$ is the spin-polarized surface density of states of the FM lead (see Appendix.~A). Clearly, the non-Hermiticity is manifested in the above system through the self-energy contribution, which lies in the lower half of the complex energy plane, leading to dissipation.

The eigenvalues of the effective Hamiltonian $\mathcal{H}_{\rm AM-FM} (\mathbf{k})$ can be obtained as 
\begin{equation}
    E^{\pm}_{\mathbf{k}} = \frac{t}{2}\lvert k \rvert^2-(2t+i\xi^{+}) \pm \alpha_{\mathbf{k}}(\lambda, \Delta, \xi^-),
    \label{eqn:d}
\end{equation}
where $\alpha_{\mathbf{k}} = \sqrt{\lambda^2\lvert k \rvert^2 + \left[\frac{\Delta}{2}\left(k_y^2 - k_x^2\right) - i\xi^{-}\right]^2}$. The condition for the appearance of degenerate points in the eigenspectrum can be written as
\begin{equation}
    \lambda^2\lvert k \rvert^2 + \frac{\Delta^2}{4}(k_y^2 - k_x^2)^2=(\xi^-)^2; \quad
    \Delta\xi^-(k_y^2 - k_x^2) = 0,
    \label{EP_Cond}
\end{equation}
resulting in the emergence of four degenerate points at $(k_x,k_y) =(\pm\frac{\xi^-}{\sqrt{2}\lambda},\pm\frac{\xi^-}{\sqrt{2}\lambda})$. These points located symmetrically with respect to the origin of $k_x-k_y$ plane. The coalescence of the eigenvectors at these points confirms the existence of four exceptional degeneracies in the AM-FM junction. We identify that the order of these EPs is \textit{two} -- characterized by the square-root energy dispersion around it and is protected by two real constraints~\cite{Sayyad_2022}. Interestingly, the parameter $\xi^{-}$ delineating a TRS-breaking perturbation (i.e., magnetization) induced by the FM lead is responsible for these two pairs of emerging second-order exceptional points. In addition, it is clear that EPs will only emerge when the magnetization $\xi^{-}$ is not orthogonal to the interface. We would like to emphasize that $d_{xy}$-wave is connected to $d_{x^2-y^2}$ via a $\pi/4$ rotation. Therefore, one can easily recognize that $d_{xy}$-AM-FM junction will also host two pairs of EPs along the axis at $(0,\pm\frac{\xi^-}{\lambda})$ and $(\pm\frac{\xi^-}{\lambda},0)$. 

\section{Topological Invariant of the EP} Now we investigate one of the topological invariants of EP, namely, the winding number which is given by~\cite{Sato_2019}
\begin{equation}
   w = \frac{1}{2\pi i}\int_{0}^{2\pi}\nabla_{\zeta}{\rm log \, det}\Phi_{\mathbf{k(\zeta)}}  \cdot d{\mathbf{\zeta}},
\end{equation}
where $\Phi_{\mathbf{k(\zeta)}} = H (\mathbf{k}) - E_{\mathbf{k}}^{\rm EP}$ around a path $\mathcal{S}^1$ encircling the EP parametrized by an angle $\zeta\in [0,2\pi]$. Interestingly, the topological charge or vorticity ($\nu$)~\cite{Nori_2017, Fu_2018} associated with an EP is connected to its winding number via the relation: $\nu=-w/2$. The difference of the eigenenergy has a square-root singularity around the EPs, which are accompanied by a branch cut that touches Riemann sheets of the complex eigenspectrum. This yields that both ${\rm EP}_1$ and ${\rm EP}_2$ have the winding number +1, therefore associated with $-1/2$ charge, whereas ${\rm EP}_3$ and ${\rm EP}_4$ have the $w=-1$ and $\nu=+1/2$.

\section{Spin projection of AM-FM Junction} To understand the behavior of spin in the AM-FM junction, we calculate the spin projection defined as $S_\alpha^{\pm} = \langle \Psi_{\pm}|\sigma_{\alpha}|\Psi_{\pm}\rangle$ where  $\alpha=x,y,z$. \textcolor{black}{The wavefunctions $\Psi_{\pm}$ can be written as
\begin{eqnarray}
&|\Psi_{\pm}\rangle = N_{\pm} \begin{pmatrix} 1 \\ f_k^{\pm} \end{pmatrix} ; |N_{\pm}|^{2} = \frac{1}{1+f_k^{\pm}f_k^{\pm *}} \nonumber \\
&f_k^{\pm} = \frac{-\frac{\Delta}{2} (k_y^2 - k_x^2) + i\xi^- \pm \alpha_k(\lambda, \Delta, \xi^{-})}{\lambda(k_y + ik_x)}, 
\end{eqnarray}
where $N_{\pm}$ is the normalization constant.
From the above equation, the spin projection along the $x (y)$ axis is given by $S_{x(y)}^{\pm} = 2|N_{\pm}|^{2}{\rm Re(Im)}[f_k^{\pm}]$. In the Hermitian phase ($\xi^{-} = 0$), the spin expectation value is obtained as $S_{x(/y)}^{\pm} = |N_{\pm}|^2 k_y(/-k_x) \frac{-\Delta (k_y^2 - k_x^2) \pm 2\alpha_k(\lambda, \Delta, 0)}{\lambda|k|^2}$. It implies that $S_{x(y)}^{\pm}$ undergoes a change of sign as we vary $k_{y(x)}$ along the diagonals $k_x = \pm k_y$ because $|N_{\pm}|^2=1/2$ under these conditions. 
However, in the NH phase ($\xi^{-} \neq 0$), the spin expectation value reduces to
\begin{equation}
    S_{x(/y)}^{\pm (\rm{NH})} \propto \frac{k_y(/-k_x)(\frac{-\Delta(k_y^2-k_x^2)}{2} \pm \alpha_k^{\rm R}) + k_{x(/y)}(\xi^- \pm \alpha_k^{\rm I})}{\lambda\lvert k \rvert^2},
\end{equation}
where $\alpha_k^{\rm{R(I)}}$ is the real (imaginary) part of $\alpha_k(\lambda, \Delta, \xi^{-})$.} The first term in the numerator represents a renormalized counterpart of the Hermitian spin projection value. On the other hand, the second term showcases a linear proportionality of the spin expectation value along the evaluated direction. It is a unique signature of non-Hermiticity, which may lead to a situation where the sign of spin projection changes with the respective momentum direction. Finally, we evaluate the spin expectation value at the EPs, which yields
\begin{equation}
  S_{x(/y)}^{\pm (\rm EP)} = \frac{k^{\rm EP}_{x(/y)} \lambda}{\xi^{-}}.
    \label{sp_EP}
\end{equation}
Remarkably, it is clear from the Eq.~(\ref{sp_EP}) that the spin expectation value coalesces at the EPs. These signatures could be used as a fingerprint of the non-Hermitian exceptional phase.

\begin{figure}
    \centering
    \includegraphics[width=1\linewidth]{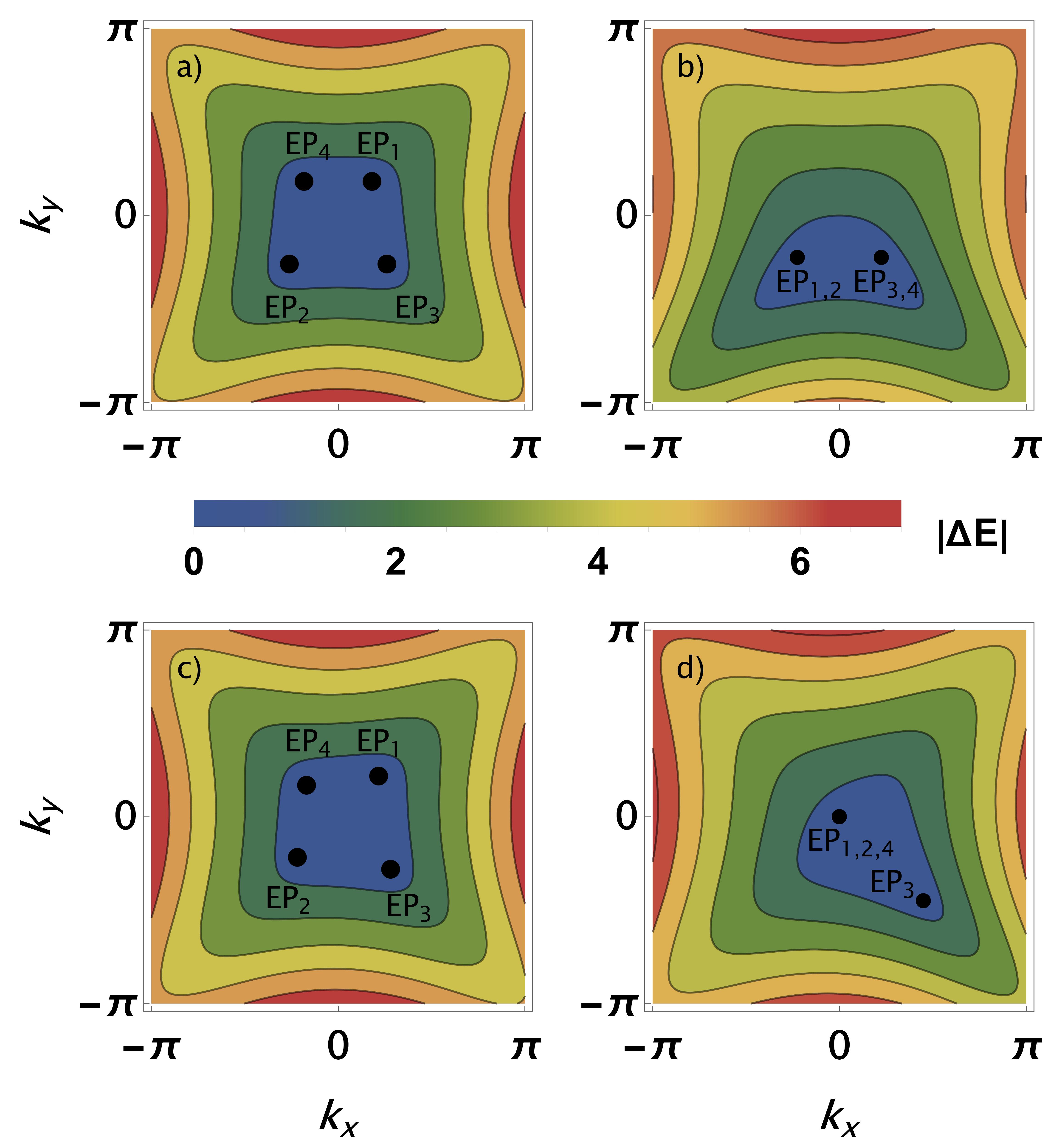}
    \caption{ The variation of the ${\rm Re}\,(\Delta E)|_{_{B\neq0}}$ has been portrayed with respect to $k_x$ and $k_y$. (a) shows four distinct EPs given by Eq.~(\ref{Eq_EP}) for $B=\xi^-/4$ and $\phi=0$. (b) reveals that with  increasing $B$, the EPs with opposite topological charges, i.e., (${\rm EP}_1$, ${\rm EP}_2$) and (${\rm EP}_3$, ${\rm EP}_4$) annihilate at $(\mp\frac{\xi^-}{\sqrt{2}\lambda},-\frac{\xi^-}{\sqrt{2}\lambda})$ for the critical field strength $B=\sqrt{2}\xi^-$ and the system is left with no EPs. (c) is similar to the scenario (a) but for $\phi = \pi/4$. (d) depicts the remarkable feature of merging three EPs, namely ${\rm EP}_1$, ${\rm EP}_2$ and ${\rm EP}_4$  at the origin for a field strength $B = \xi^-$. Increasing $B$ beyond this leads to the annihilation of two EPs, viz. ${\rm EP}_1$, ${\rm EP}_2$, and the other two EPs survive and continue their motion away from the origin. This contrasts sharply with the previous case of $\phi =0$. The other parameters used are $t=1$, $\lambda = \Delta = \xi^- = t$.}
    \label{fig:mag_tune}
\end{figure}

\section{Tuning EP at AM-FM Junction} A viable route to tune EPs is to apply an external magnetic field ($B$). The resulting Hamiltonian of the AM-FM junction in the presence of a planar magnetic field $B(\cos \phi, \sin \phi, 0)$ takes the form
\begin{equation}
    \mathcal{H}_{AM-FM}^{B}(\mathbf{k}) = \mathcal{H}_{AM-FM}(\mathbf{k}) + \mathbf{B \cdot \sigma}. \nonumber
\end{equation}
The trajectories of the EPs can be obtained by solving Eq.~(\ref{EP_Cond}) after replacing $\lambda k_y \rightarrow \lambda k_y + B \cos \phi$ and $\lambda k_x \rightarrow \lambda k_x - B \sin \phi$. The locations of EPs in the $k_x-k_y$ plane are given as: 
\begin{eqnarray}
& k^{(1)}_{\pm}=\frac{B}{2\lambda}\left[(\sin \phi - \cos \phi) \pm \sqrt{2\left(\frac{\xi^-}{B}\right)^2-(1+\sin 2\phi)}\right]   \nonumber \\
&k^{(2)}_{\pm}=\frac{B}{2\lambda}\left[(\sin \phi + \cos \phi) \pm \sqrt{2\left(\frac{\xi^-}{B}\right)^2-(1-\sin 2\phi)}\right]. \nonumber \\
\label{Eq_EP}
\end{eqnarray}
 We denote them by ${\rm EP}_1$($k^{(1)}_{+}$, $k^{(1)}_{+}$), ${\rm EP}_2$($k^{(1)}_{-}$, $k^{(1)}_{-}$), ${\rm EP}_3$($k^{(2)}_{+}$, $-k^{(2)}_{+}$) and ${\rm EP}_4$($k^{(2)}_{-}$, $-k^{(2)}_{-}$) respectively. These points lie distinctly along the diagonal on the $k_x-k_y$ plane for $\lvert B^2 \rvert < \lvert (\xi^-)^2\rvert$ as shown in fig.~\ref{fig:mag_tune}(a) and (c). 

 We now discuss the effect of the planar Zeeman field on these EPs for three special cases. {\it Case I:} $\phi = 0$ (magnetic field along the horizontal axis), with increasing $B$, ${\rm EP}_1$ and ${\rm EP}_4$ move towards the origin diagonally in the $k_x-k_y$ plane and make a crossover for $B=\xi^-$. In contrast, initially ${\rm EP}_2$ and ${\rm EP}_3$ go diagonally away from the origin but make a turnover at $B=\xi^-$. Interestingly, after the crossover, ${\rm EP}_1$ and ${\rm EP}_4$ interchange their polarity of charge. At the critical strength  $B_c=\sqrt{2}\xi^-$, both the pairs with opposite topological charges (${\rm EP}_1$,\, ${\rm EP}_2$) and (${\rm EP}_3$,\, ${\rm EP}_4$) annihilate and the system left with no EPs as shown in Fig.~\ref{fig:mag_tune}(b). {\it Case II:} $\phi = \pi/2$ (field is along the vertical axis), we observe the movement of EPs similar to $\phi=0$ case. However, in contrast to the earlier case, here ${\rm EP}_2$ and ${\rm EP}_4$ make a crossover at $B=\xi^-$ and ${\rm EP}_1$ and ${\rm EP}_3$ make a turn around. Finally, all four EPs get eliminated due to the annihilation process at $B_c$. {\it Case III:} $\phi = \pi/4$ implies the applied $B$ is in the plane with equal strength along both directions. In contrast to $\phi=0, \pi/2$, here, ${\rm EP}_1$, ${\rm EP}_2$ and ${\rm EP}_4$ proceed towards the origin diagonally, and ${\rm EP}_3$ goes away from it. Remarkably, at $B=\xi^-$, all three EPs meet at origin (see fig.~\ref{fig:mag_tune}(d)). With increasing $B$ further, ${\rm EP}_1$ and ${\rm EP}_2$ become annihilated while ${\rm EP}_3$ and ${\rm EP}_4$ continue their motion away from the origin. This indicates that the system will always have a minimum of two EPs, in contrast to the previous two cases in which the system lost all the EPs after $B_c$.

\section{Conclusions} In summary, we develop the NH theory of the $d_{x^2-y^2}$ wave altermagnet in two different scenarios. \textcolor{black}{In the first case, with the inclusion of an imaginary gauge field in the kinetic term, an intercell non-reciprocal hopping is injected into the system hosting point gap in the complex eigenspectrum without exceptional points. Interestingly, the closed nature of the eigenspectrum due to the asymmetry in the imaginary part of the energy could give rise to skin effect in AM which we leave for future study.} In the second scenario, considering an AM-FM junction, we discover the emergence of two pairs of second-order EPs having an equal magnitude of half-integer quantized topological charge with opposite polarity. Further, we confirm the existence of exceptional phase via linear variation with the respective momentum and
coalescence of the spin expectation value at the EPs. We demonstrate the tuning of the emergent EPs via a planar Zeeman field. Remarkably, by changing field direction, we reduce the number of EPs to a single pair or even null via annihilation of EPs with opposite topological charges. The prediction of emerging EPs, as well as their tunability via an external magnetic field, could make the NH altermagnet-ferromagnet junction a potential candidate for EP-enhanced sensor design. Our work provides a platform to explore the non-Hermitian physics in different kinds of collinear (such as $g$-wave and $i$-wave AMs)~\cite{Tomas_2022} and noncollinear altermagnets~\cite{Lee_2024} in future.

{\it{Note added:}} —Recently, we noticed one preprint~\cite{Narayan_2024} on the emergence of EPs in NH altermagnet-ferromagnet junction appeared in parallel with our work.

\section*{Acknowledgements}
S.N. and S.P. thank M. Sanahal for important discussions. The authors thank the anonymous referee for insightful comments.

\section*{Appendix {A}}
\label{app}
\setcounter{equation}{0}
To calculate the self-energy, we consider an open system where a two-dimensional (2D) AM represented by the Hamiltonian $\mathcal{H}^{\rm H}_{\rm{AM}}$ is coupled to a semi-infinite ferromagnetic (FM) lead denoted by the Hamiltonian $H_{\rm{FM}}$. Therefore, the open quantum system can be thought of 2D junction where the coupling at the interface is designed by the spin-independent hopping between the last site of the FM lead ($z = 0$) and the first site of the AM ($z = 1$).
The Green's function of the open quantum system is given by,
\begin{equation}
    G_{\rm AM-FM}^{r}(\omega) = \left[\omega - \mathcal{H}^{\rm L}_{\rm AM} - \Sigma^{r} (\omega)\right]^{-1}, 
\end{equation}

where $\Sigma^{r} (\omega) = V^{\dagger}g_{FM}^{r}(\omega)V$
is the self-energy in the AM due to coupling with the FM lead. Moreover, $g_{FM}^{r}$  and $V$ denote the retarded Green's function of the FM lead and the hopping matrix between FM lead and AM, respectively. Thus, we can define the effective Hamiltonian of the AM-FM junction as,

\begin{eqnarray}
    \mathcal{H}_{\rm AM-FM} = \mathcal{H}^{\rm L}_{\rm AM} + \Sigma^{r} (\omega).
\end{eqnarray}
Since the system includes an active spin degree of freedom due to FM lead, $g_{FM}^{r}$ in the spin space can be written as,
\begin{equation}\label{gf}
    g_{FM}^{r}(\omega) = \begin{pmatrix}
        g_{\uparrow \uparrow}^r &  g_{\uparrow \downarrow}^r\\
        g_{\downarrow \uparrow}^r & g_{\downarrow \downarrow}^r
    \end{pmatrix}.
\end{equation}
Now we use the recursive approach to find the Green's function of the lead $ g_{FM}^{r}(\omega)$~\cite{sp_PRR_tifm, sp_prb_scfm22, cayaoJJ, cayaosns}. We note that the FM lead contains an infinite number of lattice sites at the negative $z$-direction. Therefore, the Hamiltonian at each site of the lead is given by 
\begin{equation}
    [H_{FM}]_{i_{FM},i_{FM}} =  \epsilon_k^{FM}\sigma_0 + B \sigma_z,
\end{equation}
where $\epsilon_k^{FM} = \frac{\hbar^2}{2m}(k_x^2 + k_y^2)-\mu_{FM}$ and $B$ are the kinetic energy and Zeeman energy term in the FM lead. Here, $\mu_{FM}$ is the chemical potential in the lead. For simplicity of the notation, we define $\epsilon_{\sigma} = \epsilon_k^{FM} \pm B$ with $\sigma = \{\uparrow, \downarrow\}$ where $\pm$ represents spin-up ($\uparrow$) and spin-down ($\downarrow$) energy respectively. Therefore, the Green's function of the FM lead is given by,
\begin{equation}\label{gf2}
  [\omega-\epsilon_k^{FM} - B \sigma_z - {\cal V}^\dagger g_{FM}^r(\omega){\cal V}] g_{FM}^r(\omega) = I,
\end{equation}
where ${\cal V} = -t_z\sigma_0$ is the hopping matrix between the sites in the semi-infinite FM lead. Plugging eqn.~$\ref{gf}$ in eqn. $\ref{gf2}$, the system of equations can be obtained as
\begin{align}\label{eqn1}
    & (\omega - \epsilon_{\uparrow} - t_z^2g^{r}_{\uparrow\uparrow})g^{r}_{\uparrow\uparrow} - t_z^2g^{r}_{\uparrow\downarrow}g^{r}_{\downarrow\uparrow} = 1,\nonumber \\
    & (\omega - \epsilon_{\uparrow} - t_z^2g^{r}_{\uparrow\uparrow})g^{r}_{\uparrow\downarrow} - t_z^2g^{r}_{\uparrow\downarrow}g^{r}_{\downarrow\uparrow} = 0, \\
    & - t_z^2g^{r}_{\downarrow\uparrow}g^{r}_{\uparrow\uparrow} + (\omega - \epsilon_{\downarrow} -t_z^2g^{r}_{\downarrow\downarrow})g^{r}_{\downarrow\uparrow} = 0,\nonumber \\
    & - t_z^2g^{r}_{\downarrow\uparrow}g^{r}_{\uparrow\uparrow} + (\omega - \epsilon_{\downarrow} -t_z^2g^{r}_{\downarrow\downarrow})g^{r}_{\downarrow\downarrow} = 1.\nonumber
\end{align}
By solving eqn. $\ref{eqn1}$, we get
\begin{equation}
\label{g1}
g_{\uparrow\uparrow(\downarrow\downarrow)}^{r}(\omega) = \frac{\omega-\epsilon_{\uparrow(\downarrow)}-\sqrt{(\omega-\epsilon_{\uparrow(\downarrow)})^2-4t_z^2}}{2t_z^2}, 
g_{\uparrow\downarrow(\downarrow\uparrow)}^{r} = 0.
\end{equation}
Therefore, the retarded Green's function of the FM lead in spin basis is given by $g_{FM}^r(\omega) = {\rm diag}(g_{\uparrow\uparrow}^r,g_{\downarrow\downarrow}^r)$. It implies that the opposite spins of the lead are not coupled to each other and feel different Zeeman field. Furthermore, by analytic continuation of the frequency, we obtain
\begin{widetext}
\begin{equation}\label{case1}
    g_{\sigma\sigma}^{r}(\omega) = \begin{cases}
    \frac{1}{\lvert t_z \rvert}\left[\frac{\omega-\epsilon_\sigma}{2\lvert t_z \rvert} -{\rm sgn}(\omega-\epsilon_\sigma)\sqrt{(\frac{\omega-\epsilon_\sigma}{2\lvert t_z \rvert})^2-1}\right],& \text{if } \lvert\frac{\omega-\epsilon_\sigma}{2\lvert t_z \rvert}\rvert > 1\\
    \frac{1}{\lvert t_z \rvert}\left[\frac{\omega-\epsilon_\sigma}{2\lvert t_z \rvert} - i\sqrt{1-(\frac{\omega-\epsilon_\sigma}{2\lvert t_z \rvert})^2}\right].              & \text{if } \lvert\frac{\omega-\epsilon_\sigma}{2\lvert t_z \rvert}\rvert < 1
\end{cases}
\end{equation}
\end{widetext}
The local density of states ($\rho^{FM}(\omega) = -\frac{1}{\pi}{\rm Im}[g_{FM}^{r}(\omega)]$) of the FM lead is zero for the first case of eqn.~(\ref{case1}). On the other hand, in the second case, $g_{FM}^{r}$ hosts an imaginary term developing a finite density of states which is given by,
\begin{equation}
    \rho^{FM}_{\sigma}(\omega) = \frac{\theta(2\lvert t_z\rvert - \lvert \omega-\epsilon_{\sigma}\rvert)}{\pi\lvert  t_z\rvert}\sqrt{1-\left(\frac{\omega-\epsilon_\sigma}{2\lvert t_z \rvert}\right)^2}.
\end{equation}
Now to calculate the self-energy $\Sigma^{r}(\omega) = V^{\dagger}g^{r}_{FM}(\omega)V$, we assume $V$ is finite between the nearest-neighbor sites of the FM lead and the AM. Therefore, we can project the self-energy $\Sigma^{r}(\omega)$ to the closed AM Hamiltonian as follows
\begin{equation}
    \Sigma^{r}_{1_{\rm AM}, 1_{\rm AM}} = \langle{1_{\rm AM}}\rvert V^{\dagger}\lvert{1_{\rm FM}}\rangle \langle{1_{\rm FM}}\rvert g_{FM}^{r}\rvert {1_{\rm FM}}\rangle\langle{1_{\rm FM}}\rvert V \lvert {1_{\rm AM}}\rangle.
\end{equation}
Here, $1_{\rm FM}$ denotes the first site of the FM lead closest to the AM, and $1_{\rm AM}$ denotes the first AM site along the $z$-direction. Considering $<{1_{\rm FM}}|V|{1_{\rm AM}>} = -t'\sigma_0$ where $t^{\prime}$
is the hopping amplitude between sites 1$_{\rm FM}$ and site 1$_{\rm AM}$, we obtain
\begin{equation}
   \Sigma^{r}_{1_{\rm AM}, 1_{\rm AM}}(\omega) = \begin{pmatrix}
        |t'|^2g_{\uparrow\uparrow}^r & 0 \\
        0 & |t'|^2g_{\downarrow\downarrow}^r
    \end{pmatrix} ,
\end{equation}
where $g_{\uparrow\uparrow(\downarrow\downarrow)}^r$ is given by eqn. $\ref{g1}$. It is important to note that the imaginary part of the Green's function of the FM lead makes the self-energy non-Hermitian. 
To simplify further, we employ wideband approximation, $\lvert\frac{\omega-\epsilon_\sigma}{2 t_z}\rvert << \lvert \frac{\mu_{FM} \mp B}{2 t_z}\rvert < 1$, used in quantum transport. This allows us to neglect the dependencies of momentum and frequency of the Green's function of the FM lead. Therefore, the self-energy can be approximately written as
\begin{equation}
    \Sigma^r(\omega =0) \approx -i\xi^+\sigma_0 -i\xi^-\sigma_z,
\end{equation}
where $\xi^{\pm} = \frac{1}{2}(\xi_{\uparrow}\pm\xi_{\downarrow})$ with $\xi_{\sigma} = \pi\lvert t^{\prime}\rvert^2\rho^{FM}_{\sigma}$ and $\rho^{FM}_{\sigma}$ is given by,
\begin{equation}
    \rho^{FM}_{\uparrow(\downarrow)} = \frac{1}{t_z\pi}\sqrt{1-\left(\frac{\mu_{FM}\mp B}{2t_z}\right)^2}.
\end{equation}
\bibliography{NH_AM}
\end{document}